# Élaboration de tâches communicatives basées sur les conflits en CLE avec ChatGPT

LI Xia

Docteure en Didactique du Chinois Langue Étrangère
**Affiliation :** Université Grenoble Alpes, France.
**Courriel :** xia.li@univ-grenoble-alpes.fr
**Toile :** https://cv.hal.science/xia-li
**Adresse :** laboratoire LIDILEM, Université Grenoble Alpes, Bâtiment Stendhal, CS40700, 38058 Grenoble cedex 9, France

***Résumé :*** *Dans le cadre de l'élaboration du programme d'enseignement pour un cours d'expression orale en Chinois Langue Étrangère en milieu universitaire, l'enseignante conçoit des tâches communicatives basées sur les conflits afin d'encourager les apprenants à s'engager dans une dynamique interactionnelle et à développer leur compétence d'interaction orale. Lors de la conception de ces tâches, l'enseignante utilise ChatGPT pour l'aider à finaliser l'élaboration du programme. Cet article vise à exposer les caractéristiques saillantes des interactions entre l'enseignante et ChatGPT durant ce processus d'élaboration, ainsi qu'à examiner l'utilisation de ChatGPT et ses impacts dans ce contexte spécifique.*

***Mots clés :*** *chinois langue étrangère, intelligence artificielle, conception de programmes d'enseignement avec ChatGPT, tâche communicative basée sur les conflits*

***Title:*** *Designing conflict-based communicative tasks in Teaching Chinese as a Foreign Language with ChatGPT*

***Abstract:*** *In developing the teaching program for a course in Oral Expression in Teaching Chinese as a Foreign Language at the university level, the teacher designs communicative tasks based on conflicts to encourage learners to engage in interactive dynamics and develop their oral interaction skills. During the design of these tasks, the teacher uses ChatGPT to assist in finalizing the program. This article aims to present the key characteristics of the interactions between the teacher and ChatGPT during this program development process, as well as to examine the use of ChatGPT and its impacts in this specific context.*

# Introduction

L'avènement de l'Intelligence Artificielle (IA) ne date pas d'hier. Aujourd'hui, ses applications concernent pratiquement tous les domaines de notre vie quotidienne, y compris celui de l'enseignement/apprentissage des langues. Par exemple, l'application du Traitement Automatique des Langues (TAL) dans la traduction, tels que la traduction (comme avec DeepL[1]), la correction des erreurs linguistiques (comme Grammarly[2]) ou encore la conversion de texte en discours oral synthétisé (avec TTSmaker[3]). À la fin de l'année 2022, le lancement de l'assistance virtuelle ChatGPT[4] par OpenAI a captivé un large public, notamment grâce à ses performances remarquables en matière de traitement de l'information. Dans ce contexte, il est donc nécessaire d'examiner de plus près les applications de l'IA dans le domaine de la didactique des langues, tant sur l'enseignement que sur l'apprentissage, ainsi que les impacts qu'elles peuvent engendrer.

Cet article se concentre sur l'étude du processus d'élaboration des tâches communicatives basées sur les conflits avec ChatGPT pour un cours d'expression orale de Chinois Langue Étrangère (CLE) en milieu universitaire. Notre recherche adopte notamment une perspective descriptive (Olivier de Sardan, 2008 ; Catroux, 2018) visant à exposer les caractéristiques saillantes de l'interaction entre l'enseignante et le ChatGPT, ainsi qu'à mettre en lumière leurs impacts sur la finalisation de la conception du programme d'enseignement. Les questions de recherche sont les suivantes : comment l'utilisation de ChatGPT se manifeste-t-elle dans le processus d'élaboration des tâches communicatives basées sur les conflits ? Dans quelle mesure son utilisation influence-t-elle le programme final ?

Dans un premier temps, nous décrivons certains éléments contextuels en nous focalisant sur le processus d'élaboration de nos tâches d'études. Ensuite, nous présenterons le cadre méthodologique, comprenant le type de recherche envisagé, la constitution et le traitement du corpus, ainsi qu'une explication de certains termes de métalangage utilisés dans la partie d'analyse. Par la suite, nous nous concentrons sur le cadre théorique, qui définit certaines notions clés, telles que la tâche communicative et les intérêts d'associer la théorie du conflit dans sa conception, certains critères à prendre en compte lors de l'élaboration des tâches, ainsi que les éléments liés à l'IA. Enfin, nous exposons les résultats de l'analyse et en tirons des conclusions.

---

[1] DeepL a été lancé en 2017. Consulter le lien : https://www.deepl.com/fr/translator
[2] Grammarly a été lancé en 2009. Consulter le lien : https://www.grammarly.com/
[3] TTSmaker a été lancé en 2022. TTS signifie *Text-to-Speech*. Consulter le lien : https://ttsmaker.com/
[4] Dans notre article, le ChatGPT fait référence à la version gratuite GPT-3.5. Consulter le lien : https://chat.openai.com/

## 1. Contexte

Pour bien comprendre nos analyses, il est crucial de saisir le contexte d'application de ces tâches, les raisons de leur conception, le public cible, ainsi que le processus de leur élaboration.

### 1.1. Présentation du cours d'expression orale

Le cours d'expression orale, dispensé par l'enseignante-chercheuse (auteure de cet article) à Sorbonne Université, se déroulait une fois par semaine pendant onze semaines par semestre, avec une durée de 1 heure 30 minutes. Cet article porte sur l'étude du processus d'élaboration des tâches communicatives pour le premier cours d'expression orale du semestre 2 de l'année 2023-2024, destiné à 14 étudiants de licence 3 en Langues Étrangères Appliquées (LEA), parcours chinois. Le thème principal du cours d'expression orale s'articulait étroitement avec les autres cours, tels que la compréhension écrite, la compréhension orale et l'expression écrite, afin de maintenir la cohérence interne du *curriculum* (l'ensemble des programmes d'enseignement). Étant donné que c'était la période du Nouvel An chinois, le thème hebdomadaire se focalisait sur les phénomènes observés pendant cette période. Ainsi, le contexte des tâches communicatives se situait dans le cadre du voyage en train lors de la grande migration du Nouvel An chinois. Ensuite, celles-ci étaient réalisées en binômes.

### 1.2. Processus d'élaboration des tâches communicatives basées sur les conflits

L'élaboration de ce type de tâche a un double objectif : d'une part, placer les apprenants face à une situation conflictuelle et « fictive » (Cicurel, 2011, p. 66), ce qui leur permet de susciter des interactions orales qui se rapprochent de la situation de communication de la vie quotidienne ; d'autre part, en raison du caractère imprévu lors de la réalisation d'une telle tâche, les apprenants sont encouragés à développer une « compétence d'interaction orale », telle que postulée par Pekarek Doehler, la composante essentielle de la capacité de communication est « non pas de ce que l'apprenant est capable de produire seul, mais de ce qu'il arrive à faire avec autrui et, surtout, de ce que la situation lui demande de faire » (Pekarek Doehler, 2000, l. 26).

Nous présentons ci-dessous les différentes phases de l'élaboration de nos tâches d'étude, qui se déroulent avant, pendant et après l'utilisation de ChatGPT :

**Phase initiale - Élaboration individuelle par l'enseignante**

Cette étape initiale a pour objectif de cerner spécifiquement les sous-thèmes, à préciser les éléments contextuels et à définir les rôles A et B, le mode de déroulement, ainsi que les principaux points de conflit. Enfin, deux situations de communication sont définies (Annexe 1) : « 打电话 » (*dǎ diànhuà*, téléphoner) et « 座位 » (*zuòwèi*, siège).

**Phase intermédiaire : Élaboration avec l'utilisation de ChatGPT**

Cette phase concerne un processus de l'interaction entre l'enseignante et ChatGPT. L'enseignante recourt à ChatGPT pour deux raisons principales : la première consiste à vérifier la faisabilité et la validité des tâches conçues en demandant à ChatGPT de générer des dialogues ciblés afin d'obtenir un aperçu global des « produits finaux ». La seconde vise à solliciter ChatGPT pour fournir davantage de thèmes liés aux conflits dans ce contexte spécifique, afin d'aider l'enseignante à enrichir le programme tout en répondant aux besoins quantitatifs.

**Phase finale : Élaboration individuelle par l'enseignante**

Après avoir utilisé ChatGPT, l'enseignante s'inspire de leurs échanges pour effectuer d'éventuels ajustements visant à améliorer les tâches élaborées. Cela lui permet également d'explorer de nouvelles pistes de réflexion pour enrichir le programme final.

## 2. Méthodologie

Olivier de Sardan (2008, p. 139) souligne que la description est une forme écrite de l'observation sur le terrain de recherche. À travers d'une observation et d'une analyse minutieuse du corpus, nous pouvons percevoir, comparer, et établir des liens entre les divers phénomènes observés pour en tirer des conclusions. Comme l'indique Catroux (2018), il s'agit de « décrire des faits, des comportements par enquête, analyse de contenu, étude de cas, *etc*. […] Elle implique donc après la description, une analyse et une interprétation. Son objectif est de construire du sens ou des significations à l'aide d'un réseau d'éclairages interprétatifs […] » (Catroux, 2018, p. 13).

Notre corpus est constitué à la fois de l'interaction entre l'enseignante et ChatGPT et du programme final élaboré, afin de mettre en évidence l'impact de l'utilisation de ChatGPT sur sa finalisation. Le corpus comprend la copie de l'interface de l'interaction[5] et le programme final (Annexe 1). De plus, nous avons emprunté quelques notions clés de l'analyse conversationnelle (Bange, 1992) afin de mettre en lumière certaines caractéristiques de l'interaction entre l'enseignante et ChatGPT, ainsi que du dialogue simulé par ChatGPT. Dans un premier temps, nous segmentons les interactions sous forme de « séquences » en fonction de la nature de leurs échanges. Ensuite, nous numérotons les tours d'échanges, de la même manière que dans une « tour de parole » lors d'une conversation. Nous préférons utiliser le terme « échange » plutôt que « conversation » pour qualifier ce type d'interaction entre l'enseignante (homme) et ChatGPT (machine). Cela est dû au fait que, dans ces échanges, ChatGPT joue toujours un rôle passif tout en acceptant des instructions pour agir, ce qui le distingue des conversations humaines qui comportent d'autres caractéristiques conversationnelles telles que le « chevauchement ». Par ailleurs, pour analyser les dialogues générés en paire minimale, nous adoptons le terme « paire adjacente », comme le rappelle Bange (1992), « étant constituée de deux tours de parole en position de succession immédiate, prononcés par deux locuteurs différents et tels qu'il existe un élément reconnaissable comme le premier (*first pair part*) et un autre reconnaissable comme le second (*second pair part*) » (Bange, 1992, p. 40).

## 3. Cadre théorique

Dans cette partie, nous nous concentrons tout d'abord sur la définition de la « tâche communicative » et sa conception en lien avec la théorie du conflit socio-cognitif, ainsi que sur quelques principes à prendre en compte lors de son élaboration. Ensuite, nous nous focalisons sur la présentation du système d'IA, ses diverses composantes, ainsi que son fonctionnement.

### 3.1 Tâche communicative basée sur les conflits

Quant à la définition de la « tâche communicative », Nunan (1989) souligne sa focalisation sur le sens (*meaning-focused*) plutôt que sur la forme (*form-focused*), en la définissant comme

---

[5] Pour accéder à l'interface originale de cette interaction : https://chat.openai.com/share/118b72a5-db83-40c4-ad6e-aa180d43edc7

« *a piece of classroom work which involves learners in comprehending, manipulating, producing or interacting in the target language while their attention is principally focused on meaning rather than form* » (Nunan, 1989, p. 10). Nos tâches sont spécifiquement conçues pour résoudre des conflits afin d'encourager les apprenants à s'engager dans une dynamique interactionnelle afin de développer une « compétence d'interaction orale » (Pekarek Doehler, 2000). D'une part, cela s'explique par le fait que les conflits sont courants dans la vie quotidienne, nous avons tous vécu ou été témoins de ce type de communication. Ainsi, la réalisation de ce type de tâche revêt une valeur communicationnelle importante, ce qui permet aux apprenants de se préparer à réagir de manière appropriée face à un conflit dans un contexte similaire. D'autre part, nous nous appuyons sur la théorie du conflit socio-cognitif, tout en mettant en avant les activités cognitives qui se produisent dans la résolution des conflits, ce qui est considéré comme essentiel pour développer une compétence. Perret-Clermont (1979, p. 200) met en lumière la nécessité de la résolution du conflit pour parvenir à une adaptation intellectuelle : « […] le sujet ne pourra bénéficier de l'interaction que si elle est conflictuelle, c'est-à-dire si la diversité des centrations et la nature de la tâche de la situation exigent de sa part une réorganisation des coordinations en jeu ». De plus, Bourgeois & Nizet (1999, p. 158) soulignent que ce processus conflictuel implique notamment :

[…] une structure cognitive donnée avec une information incompatible et la perturbation cognitive qui en résulte vont engager le sujet dans la recherche d'un nouvel équilibre, recherche qui le conduira, le cas échéant, à l'élaboration d'une structure nouvelle, compatible avec l'information 'perturbante'.

Dans le cadre de l'élaboration des tâches, Ellis (2003) rappelle plusieurs critères essentiels qui influent sur leur complexité, tels que l'input, les conditions, les processus et les résultats. Étant donné que notre recherche s'inscrit dans la phase de conception des tâches, nous nous concentrons particulièrement sur les facteurs liés à l'« input », qui prend notamment en compte les éléments suivants : « (1) *medium* ; (2) *code complexity* ; (3) *cognitive complexity* ; (4) *context dependency* ; and (5) *familiarity of information* » (2003, p. 222). D'après l'auteur, le médium fait référence à la manière dont l'information est fournie, par exemple sous forme orale ou écrite. La complexité du code se rapporte aux éléments linguistiques utilisés, tels que la structure syntaxique et le choix du lexique. Quant à la complexité cognitive, elle concerne le processus impliquant les types et la quantité d'informations à traiter pour accéder au sens. La dépendance au contexte concerne les supports utilisés pour accéder aux informations, tels que les images ou les tableaux. Il est également important de tenir compte de la familiarité de l'information, qui fait référence à « *the relationship between the theme of the task and the individual learner's knowledge* » (Ellis, 2003, p. 223). Il est à noter que, durant ce processus d'élaboration, l'enseignant joue un rôle crucial en tenant compte de ces critères afin d'optimiser le processus d'enseignement/apprentissage.

## 3.2 Intelligence artificielle

En ce qui concerne l'interprétation de l'IA, Herbert (1996/1969), l'un des pionniers dans ce domaine, met en avant le fait que l'IA ne s'oppose pas aux « sciences naturelles », car elle n'est pas conçue pour violer les lois naturelles. Au contraire, elle est constamment en adaptation pour répondre aux besoins et aux objectifs de l'humain. Ainsi, l'auteur préfère qualifier le terme d'« artificiel » en l'associant à l'« art » et spécifie quatre aspects pour saisir son sens :

*1. Artificial things are synthesized (though not always or usually with full forethought) by human beings. 2. Artificial things may imitate appearances in natural things while lacking, in one or many respects, the reality of the latter. 3. Artificial things can be characterized in terms of functions, goals, adaptation. 4. Artificial things are often discussed, particularly when they are being designed, in terms of imperatives as well as descriptives* (Herbert, 1996/1969, p. 5).

Frécon & Kazar (2009, p. 1) proposent une notion plus restreinte de l'intelligence qui se focalise notamment sur la capacité de « résolution de problème », les auteurs classifient quatre niveaux de manifestations de l'intelligence : la compréhension d'une situation, l'application d'un cadre de raisonnement fourni dans la résolution d'un problème, la vérification d'un système de raisonnement établi ou la création d'un nouveau cadre, ainsi que la résolution de problèmes dans des conditions en constante évolution.

Selon Tisseau (2020/1996), l'IA est « un système possède une certaine structure, et il est parcouru par des flux de matière, d'énergie et d'information. Il est soumis à des mécanismes internes qui lui permettent de remplir certaines fonctions » (Tisseau, 2020/1996, p. 26). Cela signifie que L'IA comme l'intelligence humaine est un système général qui se compose de plusieurs systèmes assumant respectivement de diverses fonctions pour assurer la performance du sujet. D'après l'auteur, ce qui distingue l'IA des autres programmes informatiques classiques, c'est son système heuristique, qui met l'accent sur la capacité d'apprentissage et de découverte. Tout comme chez l'être humain, l'IA dispose également de deux types de connaissances : les connaissances déclaratives et les connaissances procédurales. Comme le précise l'auteur, « une partie de leurs connaissances est stockée sous forme explicite, comme dans un livre, et une autre partie sous forme de programmes capables d'effectuer certaines tâches » (Tisseau, 2020/1996, p. 94). Cependant, malgré le fait que le système d'IA dispose d'une grande quantité de connaissances dans ces deux types, il rencontre encore des contraintes dans l'assimilation de certaines connaissances « non dites » et « non écrites », comme l'illustré l'auteur : « mais même si toutes les connaissances étaient formulées dans des livres, il ne suffirait pas de les transférer directement dans les ordinateurs pour que ceux-ci les comprennent et sachent les utiliser » (Tisseau, 2020/1996, p. 97). En ce sens, Diallo (2023) affirme que ChatGPT n'est pas intelligent et qu'il est encore loin d'être créatif :

Il n'est qu'un outil capable de rassembler en un temps record des données ciblées et d'en faire une synthèse selon la fréquence de leur apparition sur le web. Il s'appuie sur l'opinion la plus répandue (mainstream) et passe à côté des nuances, des opinions minoritaires et des controverses scientifiques (Diallo, 2023, p. 10).

Ce point de vue se confirme dans l'analyse suivante de la manière dont ChatGPT génère des dialogues.

## 4. Analyse

Dans cette partie, nous mettrons en évidence les principaux aspects de l'interaction entre l'enseignante et ChatGPT, tout en soulignant les rôles joués par ChatGPT dans ce processus de l'élaboration de nos tâches d'étude, ainsi que leur impact sur l'achèvement du programme final. Une analyse du dialogue généré par ChatGPT sera également présentée pour attirer notre attention sur certains points à considérer lorsqu'on souhaite utiliser ces dialogues à des fins pédagogiques.

### 4.1. Évaluation de la faisabilité de la tâche avec ChatGPT

La première séquence interactionnelle entre l'enseignante et ChatGPT commence par une demande explicite de l'enseignante. Elle présente les deux tâches qu'elle a initialement conçues, en fournissant les consignes (élaborer un dialogue basé sur les situations de communication données), le contexte (dans le train lors de la migration du Nouvel An chinois), les rôles (A & B) et les sujets de conflit. En ce qui concerne la forme de présentation, l'enseignante adopte un style structuré, en numérotant les deux scènes, en spécifiant les rôles A et B, tout en utilisant une ponctuation appropriée. ChatGPT respecte les consignes de l'enseignante et génère deux dialogues, chacun comprenant généralement trois paires adjacentes :

**Enseignante 1 (E1)**

根据下面情景编对话： 背景：在春运的火车上。
1. 打电话。
A: 正在车厢里打电话，而且声音特别大。
B: 声音打扰到他休息，希望对方声音小一点，或者去过道里打。
2. 座位。
A 希望和她的朋友坐在一起，所以想和你换座位，但是你不想换，因为她的座位不靠窗户，你想坐在靠窗的位置。

**ChatGPT 2 (C2)**
情景 1：在火车车厢里打电话
A1：喂，你好，我在车厢里，你能听见吗？
B2：能听见，但你的声音太大了，有点吵到我休息了。
A3：哦，抱歉啊，我没注意到。这样好一点吗？
B4：嗯，小点声吧，或者你能不能到过道里打电话？
A5：过道里可能信号不太好，我就在这里打一下，小点声。
B6：谢谢，那就好。

情景 2：换座位问题
A1：嗨，你好，我是在旁边的座位上的，不知道能不能和你换个座位？
B2：哦，你好。为什么要换座位呢？
A3：我和我的朋友一起来的，她在车厢另一头，我们想坐在一起聊天。
B4：噢，抱歉，但是我比较喜欢坐在靠窗的位置，这个座位离窗户挺近的。
A5：哦，我理解。可是我朋友更喜欢靠窗的位置，你能不能让一下呢？
B6：抱歉啊，我真的喜欢这个座位，离窗户近，风景好。能不能找其他人换呢？
A7：好吧，我看看还有没有其他人愿意换。谢谢你的理解。

Au niveau de la forme, la réponse de ChatGPT (C2) présente un style bien élaboré, voire amélioré, avec l'ajout du titre « 情景 » (*qíngjǐng*, situation de communication) et un enrichissement de celui-ci, par exemple, « 在火车车厢里打电话 » (*zài huǒchē chēxiāng lǐ dǎ diànhuà*, téléphoner dans le wagon du train ) plutôt que simplement « 打电话 » (*dǎ diànhuà*). Sur le plan du contenu, dans le premier dialogue, il apporte de nouvelles informations, notamment la raison pour laquelle A ne souhaite pas sortir dans le couloir pour passer son appel en raison d'une mauvaise réception du signal. Pour le deuxième dialogue, ChatGPT fournit également des détails supplémentaires en précisant la position de A et celle de son amie, située à l'autre bout du wagon, ainsi que la raison pour laquelle A souhaite changer de place pour discuter avec son amie, tandis que B préfère ne pas changer de siège pour pouvoir admirer le paysage à travers la fenêtre.

Ces propositions de dialogues simulés permettront à l'enseignante d'envisager la structure du dialogue que les apprenants pourraient réaliser, ainsi que les informations qui pourraient y être abordées. L'élaboration de ce programme étant à un stade préliminaire, il est essentiel pour l'enseignante de pouvoir vérifier la faisabilité de la tâche afin d'en garantir la pertinence. Cela permet ensuite aux apprenants d'initier leur propre dialogue sur le sujet proposé.

Il convient maintenant d'examiner quelques caractéristiques de ces dialogues générées par ChatGPT. Dans le premier dialogue, nous remarquons une confusion de l'interlocuteur dans la première paire adjacente. Lorsque A passe l'appel en demandant au destinataire de l'appel « 你能听见吗？» (*Nǐ néng tīngjiàn ma* ? Tu m'entends ?), c'est B qui fournit une réponse affirmative « 能听见 » (*néng tīngjiàn*, oui*)*, puis B exprime son mécontentement, ce qui montre une incohérence discursive. Dans le deuxième dialogue, en A5, A tente d'imposer sa demande à B en exprimant directement « 你能不能让一下呢？» (*Nǐ néng bu néng ràng yíxià ne* ? Peux-tu céder la place ou pas ?)*, ce qui montre un caractère agressif et ne correspond pas aux principes de communication.

Par ailleurs, nous observons que, afin de renforcer les caractéristiques d'oralité au sein des dialogues, ChatGPT s'efforce d'utiliser de manière rigide des régulateurs, ce qui n'est parfois pas du tout adapté à la situation de communication, tels que « 哦 » (*ò*, comme « ok » ou « oui » en français pour exprimer un accord ou manifester une attitude sceptique), « 嗯 » (*èn*, comme « hm » en français pour indiquer un accord), « 噢 » (*ō*, utilisé pour indiquer un moment où l'on réalise soudainement quelque chose ou exprimer un certain d'étonnement). Par exemple, dans le deuxième dialogue, en B2, B répond à la salutation de A par « 你好 » (*nǐhǎo*, bonjour), mais il est précédé d'un régulateur « 哦 » (*ò*), au lieu de lui répondre directement par « 你好 ». De plus, concernant la forme linguistique, il arrive parfois que des formulations inhabituelles, voire agrammaticales et redondantes, se présentent comme en A1 dans la deuxième situation de communication, avec « 我是在旁边的座位上的 » (*Wǒ shì zài pángbiān de zuòwèi shàng de*), au lieu d'une formulation plus concise et précise, telle que « 我坐在旁边 » (*Wǒ zuòzài pángbiān,* je suis assise à côté).

Après avoir présenté certaines caractéristiques essentielles des dialogues générés par ChatGPT, nous pouvons maintenant revenir à l'objectif principal de l'enseignante, qui cherche avant tout à obtenir une vue d'ensemble du « produit final ». Étant donné que le dialogue généré n'est pas assez long, l'enseignante (E3) poursuit sa demande en insistant sur la nécessité de se concentrer sur les conflits :

**Enseignante 3 (E3)**
如果结果是冲突呢
**ChatGPT 4 (C4)**
**情景1：在火车车厢里打电话**
A1：喂，你好，我在车厢里，你能听见吗？
B2：能听见，但你的声音太大了，有点吵到我休息了。
A3：哦，抱歉啊，我没注意到。这样好一点吗？
B4：小点声吧，或者你能不能到过道里打电话？

A5：过道里可能信号不太好，我就在这里打一下，小点声。
B6：你就不能注意一下吗？真的很影响休息。
A7：我也是有事情要处理，没办法，你能不能忍一下？
B8：唉，好吧，但尽量别太吵。

**情景 2：换座位问题**
A1：嗨，你好，我是在旁边的座位上的，不知道能不能和你换个座位？
B2：哦，你好。为什么要换座位呢？
A3：我和我的朋友一起来的，她在车厢另一头，我们想坐在一起聊天。
B4：噢，抱歉，但是我比较喜欢坐在靠窗的位置，这个座位离窗户挺近的。
A5：可是我朋友更喜欢靠窗的位置，你能不能让一下呢？
B6：不行，我就是喜欢这个座位。找别人换吧。
A7：真的没有其他办法吗？我们真的很想在一起坐。
B8：对不起，我不想换座位，你们找别的解决办法吧。

在冲突的情况下，双方可能会产生一些不满和矛盾，但也可以通过理性沟通和尊重对方的意见来寻找解决方案。

Dans le premier dialogue, en se basant sur les dialogues déjà générés en C2, ChatGPT rajoute une paire adjacente, comme illustré en B6 et A7. Après avoir montré les efforts que A5 va faire en parlant un peu moins fort, en B6, il insiste sur le fait qu'il doit faire attention, car cela influence réellement son repos. A7 répond qu'il a des affaires à traiter, donc il n'a pas le choix et termine par « 你能不能忍一下？ » (*Nǐ néng bu néng rěn yíxià ?* Peux-tu tolérer un peu ou pas ?). Le style discursif de cette paire montre son agressivité. Ensuite, en B8, le locuteur marque son acceptation. Durant ce processus, nous constatons fortement un caractère « artificiel » qui ne correspond pas à la manière habituelle de mener une conversation, en raison de son style discursif et de la façon dont la thématique se développe. Par ailleurs, il est à noter qu'à la fin de son tour, ChatGPT identifie qu'il s'agit d'une situation conflictuelle qui risque de violer ses principes éthiques. Par conséquent, il souligne les attitudes encouragées face aux conflits afin de les éviter autant que possible.

Bien que la qualité du dialogue soit médiocre, celui-ci illustre néanmoins comment une conversation conflictuelle pourrait se dérouler, en mettant en évidence ses points de tension et les structures discursives que les apprenants pourront développer lors de la phase de réalisation des tâches. Cela répond à la demande initiale de l'enseignante, qui souhaitait obtenir un aperçu global du déroulement de ce type de conversation afin d'évaluer la faisabilité et la pertinence des tâches proposées.

# 4.2. ChatGPT : un partenaire pour la contribution de nouvelles connaissances et une source d'inspiration

L'enseignante a besoin de 7 sujets qui traitent des points conflictuels dans le contexte du train, afin de les attribuer aux 7 groupes de binômes, ce qui constitue une demande quantifiée d'exemples. Ainsi, dans la suite de l'échange, en E5, elle demande explicitement à ChatGPT de fournir d'autres exemples. En réponse, ChatGPT (C6) adopte le même modèle de présentation, numérote les différentes scènes et respecte scrupuleusement les consignes initiales pour générer des dialogues :

**Enseignante 5 (E5)**
还有什么其它冲突性的例子在火车上
**ChatGPT 6 (C6)**
情景 3：音量冲突
A1：你好，能不能把音乐的声音调小一点？太大声了。
B2：哦，抱歉啊，我没注意。这样可以吗？
A3：还是有点大，能不能再小一些？
B4：好吧，这样可以了吗？
A5：嗯，好了，谢谢。
情景 4：空间冲突
A1：不好意思，能否把你的行李往一边移一下？我的包放不下了。
B2：啊，这里挤不下了。你看能不能找别的地方放？
A3：我找了很久了，这是唯一一个位置了。你能不能挪一下？
B4：我也没地方放，你看你能不能把行李塞到其他地方？
A5：好吧，我再找找看。
在火车上，由于空间有限和乘客众多，有时候会发生音量、座位和行李等方面的冲突。解决这些冲突的关键是通过礼貌的沟通，尽量达成双方都能接受的妥协。

Nous remarquons que ChatGPT propose deux scènes conflictuelles : l'une concerne le volume de la musique, tandis que l'autre aborde le problème d'espace lié au rangement des valises. À la fin de l'échange, il résume brièvement les points essentiels des conflits potentiels dans le train et mentionne également d'autres aspects, tels que le problème de siège. De plus, il rappelle systématiquement les règles éthiques à respecter pour éviter les conflits.

À partir de la réponse (C6), l'enseignante filtre les informations et sélectionne celles qui correspondent à ses objectifs d'enseignement/apprentissage. Vu que le sujet de volume liée à la musique est similaire au premier sujet de la tâche concernant le volume de la conversation téléphonique dans le train, l'enseignante l'élimine afin d'éviter une répétition de sujets. Toutefois, le conflit lié au rangement des valises retient particulièrement l'attention de l'enseignante, ce qui représente une nouvelle contribution dans l'élaboration de son programme. Ainsi, en E7, l'enseignante souhaite voir le déroulement de ce dialogue et lui demande simplement « problème de valise » :

**Enseignante 7 (E7)**
3. 行李问题。
**ChatGPT 8 (C8)**
情景3：行李冲突
A1: 你好，不好意思，能麻烦你往一边挪挪行李吗？我的行李放不下了。
B2: 啊，这里真的很挤，我也没法再往一边移了。
A3: 我理解，可是我这边也没有别的地方可以放行李了。
B4: 唉，也是，这火车上的空间确实有限。
A5: 你看能不能调整一下，也让其他人方便一些？
B6: 我也是想找个舒服点的位置放行李，不过这里真的不好找。
A7: 对不起，我也不是有意找麻烦的，只是实在没地方放了。
B8: 行了行了，我挪挪看，但是这是有限度的。
在行李问题上，很容易因为空间有限而引起冲突。尽量通过理解对方的立场，共同探讨解决方案，避免产生过多的紧张气氛。

En réponse, ChatGPT développe un dialogue relativement plus long, qui comprend quatre paires adjacentes, contre deux dans les dialogues précédents. Malgré sa longueur, le contenu présente toujours une certaine « redondance » dans le développement thématique, ainsi que des usages inappropriés et inhabituels dans le style discursif et les formes linguistiques, comme mentionné précédemment. Ensuite, en comparant avec le programme final, nous remarquons que le sujet 3 de la tâche s'inspire du thème de l'exemple concernant le problème des valises. Néanmoins, l'enseignante a modifié le point conflictuel au lieu de suivre la proposition de ChatGPT, qui présente A demandant à B de lui accorder plus de place pour ranger ses valises. Elle a précisé que A a trop de valises et occupe trop d'espace, ce qui rend B inconfortable, comme illustré ci-dessous :

**Programme final (Annexe 1)**

3.　　　　行李。A 的行李很多，占了你的座位空间，挤得你很不舒服。

Cette modification de contenu résulte du fait que l'expérience personnelle de l'enseignante dans le train, qui révèle une certaine fréquence de conflits, conduit à une évaluation selon laquelle sa valeur communicationnelle est plus élevée.

Par la suite, l'enseignante (E9) continue de solliciter ChatGPT pour qu'il fournisse d'autres exemples du même type. Dans ce contexte, ChatGPT propose cette fois-ci cinq thèmes. Cela montre qu'une consigne d'élicitation tout au long de l'échange est nécessaire pour obtenir une production plus variée. Les sujets proposés par ChatGPT portent sur : l'ajustement du siège, l'odeur, la température, le volume des conversations et l'occupation des places assises. À partir de cette réponse, nous constatons que dans le programme final de l'enseignante, elle a conservé les sujets de l'ajustement du siège et de l'odeur.

**Enseignante 9 (E9)**
还有其他在火车上容易发生冲突的例子吗?
**ChatGPT 10 (C10)**
当在火车上有限的空间里,乘客们可能会面临各种潜在冲突。以下是一些可能发生的情景:
座位倾向冲突:
A:"嗨,你能不能不要把腿伸得那么远?我的腿都没地方放了。"
B:"我也没办法,座位太挤了,我只能这样坐。"
气味冲突:
A:"不好意思,你能不能不要吃这么味道重的东西吗?有点受不了。"
B:"哦,真的吗?抱歉,我没想到会影响到别人。"
温度冲突:
A:"车里怎么这么冷啊,你能不能不要打开窗户?"
B:"我反而觉得有点闷,开窗户通通风好吗?"
对话声音冲突:
A:"你们说话能小声点吗?我在这边睡觉呢。"
B:"抱歉,我们只是想聊聊天,没想到会吵到你。"
座位争夺冲突:
A:"这是我的座位,你怎么可以坐在这里?"
B:"哦,不好意思,我看到这里没人就坐下了,你没放个东西占位啊。"
在这些情况下,理性沟通和互相尊重是解决冲突的关键。尽量以礼貌的方式表达自己的需求,同时也要考虑到其他人的感受和需求。

**Programme final (Annexe 1)**

5. 调椅背。A 坐在你前面,他希望自己能有更多空间,所以他把椅背向后调了很多,挤得你没地方了。
6. 吃东西。A 在吃味道很大的东西,让你实在受不了。

Concernant le sujet de l'odeur, elle a conservé le même point conflictuel, qui est considéré comme courant dans la vie quotidienne. Quant au conflit du siège, la proposition de ChatGPT a rappelé à l'enseignante des situations similaires qu'elle a rencontrées dans sa propre vie quotidienne : certaines personnes sont dérangées par celles qui s'assoient devant et inclinent leur siège sans demander l'avis de l'autre, un scénario qui se produit plus fréquemment que celui proposé par ChatGPT. Rappelons que la conception de ce type de tâche s'inscrit dans une perspective communicationnelle, l'enseignante souhaite encourager les apprenants à s'exprimer davantage à travers cette tâche. Ainsi, il est essentiel de respecter le principe de familiarité dans sa conception. Durant ce processus, nous observons que l'enseignante s'inspire initialement des exemples donnés par ChatGPT. Ce dernier lui permet de faire un rappel de mémoire des éléments liés, puis elle évalue la pertinence de ces exemples en fonction de ses objectifs didactiques préétablis et de ses connaissances générales. Enfin, elle crée le sujet 5 qui lui parait plus approprié et bénéfique pour les apprenants. L'accompagnement de ChatGPT permet à l'enseignante de rappeler progressivement des témoignages relatifs et de créer les deux autres sujets de manière indépendante : l'un concerne la peur de l'animal de compagnie, et l'autre concerne des bruits générés par un groupe de jeunes jouant aux cartes :

**Programme final (Annexe 1)**

4. 　　　宠物。A 带着他的宠物狗坐火车。可是，你很害怕狗。
7. 　　　玩牌。一群学生在玩牌，他们一边玩一边聊天，声音特别大，影响到你休息了。

En effet, la plupart de ces scènes conflictuelles sont issues de véritables témoignages tirés de l'expérience personnelle de l'enseignante. Avec l'utilisation de ChatGPT, elle a pu se rappeler ces moments liés et évaluer leur pertinence dans la mise en œuvre de ces tâches, ce qui lui a permis également de gagner en efficacité dans son travail d'élaboration.

Dans la poursuite de l'échange, l'enseignante souhaite obtenir un aperçu global du déroulement du dialogue concernant le sujet 7. Ainsi, en E11, elle sollicite explicitement ChatGPT en lui donnant une consigne claire : « développer un long dialogue » en suivant la scène de communication proposée :

**Enseignante 11 (E11)**
发展一个长对话：7. 　　　玩牌。一群学生在玩牌，他们一边玩一边聊天，声音特别大，影响到你休息了。
**ChatGPT 12 (C12)**
A1: 喂，大家好！这把扑克太刺激了！
B2: 嗯，没错，这局真是精彩。你看我这手牌...
C3: 哈哈，别说了，我手气一直不好。不过今天大家真是热闹！
A4: 对啊，这种时候聚在一起玩牌最开心了！
D5: 不好意思，打扰一下。能不能小声点？我正在休息呢。
A6: 哦，抱歉啊，没有注意到。我们就是兴奋地玩牌，影响到你了吗？
D7: 是啊，声音太大了，很影响休息。
B8: 不好意思，我们确实有点太 high 了。要不我们稍微调整一下音量？
D9: 谢谢，那就好。
C10: 没事，没事，我们以后小声点。继续玩吧！
A11: 哈哈，好的，继续！这把我可是要赢回来。
在这个对话中，D 以一种礼貌的方式提醒其他人他正在休息，而 A、B 和 C 也都表示了理解，并愿意调整音量，以维护和谐的氛围。这样的沟通可以有效地解决问题，保持友好关系。
**Enseignante 13 (E13)**
更长一点的
**ChatGPT 14 (C14)**
A1: 大家好！今天真是个好日子，玩牌的氛围特别好啊！
B2: 是啊，这把斗志昂扬，谁也不肯轻易认输！
C3: 哈哈，就是，这群人都超有趣，让人忍不住狂笑。

A4: 我们这是打得不亦乐乎啊！这是我这个月最开心的一天了。
D5: 不好意思，打扰一下。你们玩得太 high 了，能不能稍微小声一点？我正想休息呢。
A6: 哦，抱歉啊，没有注意到。我们确实声音大了点。
B7: 哎呀，真是不好意思，没想到会影响到你。我们稍微收敛一下声音。
D8: 谢谢，我真的需要休息一下。你们可以继续玩，只是尽量别太吵就好。
C9: 没问题，我们知道了。你先休息，我们悄悄地玩。
A10: 当然，我们会注意的。休息好了别忘了回来，我们还有很多好玩的牌局。
D11: 好的，谢谢你们的理解。
在这个对话中，D 表达了自己的需求，而其他人则以友好的态度回应，并表示会调整音量，保证不再打扰到 D 的休息。这种沟通方式有助于维护团体关系的和谐。

Afin de prolonger la longueur du dialogue, nous constatons que ChatGPT crée quatre rôles tout en respectant le contexte d'un groupe d'étudiants. Le premier dialogue comprend au total 11 tours de parole. Par la suite, l'enseignante (E13) lui demande de développer un dialogue encore plus long. Cependant, ChatGPT n'a ni étendu le nombre de tours de parole, ni ajouté de nouvelles informations pour enrichir la thématique du dialogue. En revanche, il introduit des mots plus formels, tels que « 斗志昂扬 » (*dòuzhì ángyáng*, plein de combativité), « 不亦乐乎 » (*bú yì lè hū*, s'amuser énormément), « 收敛 » (*shōuliǎn*, se retenir), *etc*. Cela donne un style discursif moins naturel et réduit l'oralité du dialogue. De plus, comme le montrent A10 et D11, bien que D ne fasse pas partie de leur groupe, il intervient en tant que voisin pour exprimer son mécontentement, tandis qu'A l'invite à jouer après s'être bien reposé. Cela entraîne une confusion concernant le destinataire et crée une incohérence logique dans le développement de la conversation.

## Conclusion

À travers cette analyse, on aperçoit que l'enseignante occupe une position centrale en tant que « commandante » et « juge » au cours de l'interaction. Parmi les exemples de mots sélectionnés pour illustrer ses activités cognitives figurent : demander, filtrer, sélectionner, modifier, conserver, s'inspirer de, évaluer, créer, *etc*. En revanche, le ChatGPT assume notamment un rôle de partenaire de contribution et de source d'inspiration pour aider l'enseignante à évaluer et enrichir le programme, notamment en répondant à des demandes quantitatives. Cela lui permet de gagner en efficacité dans la finalisation de l'élaboration du programme d'enseignement. En ce qui concerne le fond du sujet proposé, il est crucial de toujours évaluer sa valeur communicationnelle (par exemple, cela peut-il refléter les intentions des locuteurs ? Est-il courant dans la vie quotidienne ?) ainsi que sa valeur didactique (correspond-il à nos objectifs d'apprentissage ?). Les modifications apportées entre ce que ChatGPT a proposé et ce que l'enseignante a retenu dans le programme final reflètent ce processus de réflexion.

Comme nous l'avons constaté, la qualité du dialogue généré présente plusieurs problèmes, notamment des contradictions et des usages linguistiques inappropriés. Cela découle du fait que le système d'IA ne possède pas la capacité optimale pour détecter les connaissances implicites, ce qui entraîne parfois des résultats insatisfaisants. Dans notre programme d'étude, le dialogue généré par ChatGPT remplit principalement une fonction « informative ». Néanmoins, il peut également servir de « modèle » dans la phase post-tâche, lorsque des retours linguistiques sont fournis aux apprenants sur leurs réalisations. À ce moment-là, il est également pertinent de leur présenter ces dialogues afin de mettre en évidence des expressions pertinentes ou des points argumentatifs nouveaux. C'est pourquoi il est essentiel pour les enseignants d'adopter une attitude prudente et de veiller à la qualité des dialogues générés lorsqu'ils sont utilisés comme modèles à des fins pédagogiques.

## Annexe 1 Programme final

**两两一组，选择下面的主题之一，准备一个三分钟的对话。背景：在春运的火车上。**

1. 打电话。A 打电话的声音特别大，打扰到了你休息。
2. 座位。A 希望和她的朋友坐在一起，所以想和你换座位，但是你不想换。
3. 行李。A 的行李很多，占了你的座位空间，挤得你很不舒服。
4. 宠物。A 带着他的宠物狗坐火车。可是,你很害怕狗。
5. 调椅背。A 坐在你前面，他希望自己能有更多空间，所以他把椅背向后调了很多，挤得你没地方了。
6. 吃东西。A 在吃味道很大的东西，让你实在受不了。
7. 玩牌。一群学生在玩牌，他们一边玩一边聊天，声音特别大，影响到你休息了。

## Références